\documentclass[lettersize,journal]{IEEEtran}
\usepackage{cite}
\usepackage{amsmath,amssymb,amsfonts}
\usepackage{algorithmic}
\usepackage{graphicx}
\usepackage{subfigure}
\usepackage{subcaption}
\usepackage{textcomp}
\usepackage{xcolor}
\usepackage{booktabs}
\usepackage[ruled,linesnumbered]{algorithm2e}
\usepackage{float}
\usepackage{hyperref}
\hypersetup{
hypertex=true,
colorlinks=true,
linkcolor=blue,
anchorcolor=blue,
citecolor=blue
}
\begin{document}
\let\clearpage\relax
\title{CAPSim: A Fast CPU Performance Simulator \\Using Attention-based Predictor}
\author{
Buqing Xu\textsuperscript{1},
Jianfeng Zhu\textsuperscript{1},
Yichi Zhang\textsuperscript{1},
Qinyi Cai\textsuperscript{1},
Guanhua Li\textsuperscript{1},
Shaojun Wei\textsuperscript{1},
Leibo Liu\textsuperscript{1}\\
\textit{\textsuperscript{1}School of Integrated Circuits, BNRist, Tsinghua University, Beijing, China}\\
\texttt{\{xbq22, zhangyc22, cqy24\}@mails.tsinghua.edu.cn,\\
\{zhujianfeng, ligh429, wsj, liulb\}@tsinghua.edu.cn}
\thanks{This work was supported in part by the National Natural Science Foundation of China (Grant  No. 62204139, No. U22B2024), and in part by No.31513010104.}
}

\maketitle
\thispagestyle{empty}
\begin{abstract}
CPU simulators are vital for computer architecture research, primarily for estimating performance under different programs.
This poses challenges for fast and accurate simulation of modern CPUs, especially in multi-core systems.
Modern CPU peformance simulators such as GEM5 adopt the cycle-accurate and event-driven approach, which is timeconsuming to simulate the extensive microarchitectural behavior of a real benchmark running on out-of-order CPUs. Recently, machine leaning based approach has been proposed to improve simulation speed, but they are currently limited to estimating the cycles of basic blocks  rather than the complete benchmark program. 

This paper introduces a novel ML-based CPU simulator named CAPSim, which uses an attention-based neural network performance predictor and instruction trace sampling method annotated with context.
The attention mechanism effectively captures long-range influence within the instruction trace, emphasizing critical context information. This allows the model to improve performance prediction accuracy by focusing on important code instruction.
CAPSim can predict the execution time of unseen benchmarks at a significantly fast speed compared with an accurate O3 simulator built with gem5. Our evaluation on a commercial Intel Xeon CPU demonstrates that CAPSim achieves a $2.2-8.3\times$ speedup compared to using gem5 built simulator,
which is superior to the cutting-edge deep learning approach.
\end{abstract}


\section{Introduction}
\IEEEPARstart{C}{omputer} architecture researchers demand fast and accurate performance simulators to study complicated architecture design and software optimization~\cite{hennessy2011computer}. 
Currently, there exist simulators that are cycle-accurate, event-driven, or interval-based, etc. Each with varying degrees of speed and precision. These different simulators are tailored to distinct application scenarios and functions. For example, compiler designers often need a performance simulator at the basic block level, while software or hardware developers require a cycle-accurate simulator that supports complete code-level execution
\cite{llvmmca},
\cite{abel2022uica},
\cite{Intel2017},
\cite{laukemann2019automatic},
\cite{laukemann2018automated}.
which can provide engineers with insights into potential bottlenecks within such systems~\cite{DBLP:conf/sc/CarlsonHE11}, \cite{loh2009zesto}, \cite{DBLP:conf/dac/PatelACG11}, \cite{binkert2011gem5}.

With the advancement of CPU microarchitecture and the emergence of multi-core systems, the simulation time of processor simulators has become increasingly prolonged, which has turned into an obstacle to the development of computer architecture. Low simulation speed is often attributed to two main reasons: (1) the cycle-accurate simulation of running a standard benchmark, e.g., SPEC 2017 suite, is mostly a sequential process that is difficult to be accelerated by parallel methods like multithreading, (2) simulating a modern out-of-order CPU requires prohibitive time caused by modeling all microarchitecture details.
Besides, the challenge of inaccuracy remains a common concern when utilizing a cycle-level simulator. Inaccuracy often arises from discrepancies between the behaviors of the simulator and the actual processor, stemming from the complexity of the real processor. Intricate details of their microarchitecture designs are also challenging to abstract or confidential, rendering accurate simulation a difficult task. 

Numerous studies try to use machine learning (ML) approaches to increase the speed and accuracy of performance prediction. With the advancement of artificial neural networks
\cite{DBLP:journals/corr/GuWKMSSLWW15},
\cite{DBLP:journals/tnn/WuPCLZY21},
\cite{DBLP:journals/corr/abs-1801-01078},
\cite{AttentionIsAllYouNeed},
a plethora of data-driven approaches have emerged for conducting performance prediction.  Ithemal \cite{DBLP:conf/icml/MendisRAC19} uses the Long Short-Term Memory (LSTM) neural network to predict the throughput of basic blocks. Difftune \cite{DBLP:conf/micro/RendaCMC20} leverages the Ithemal network to optimize the hardware parameters embedded in the llvm-mca simulator~\cite{llvmmca}, resulting in higher prediction accuracy. Granite \cite{DBLP:conf/iiswc/SykoraPMY22} develops a graph neural network to predict the basic block throughput, introducing a novel perspective on the matter. Besides of these, traditional ML algorithms, such as the regression model, is also used to predict the cycle per instruction (CPI) of O3 superscalar processors, along with the power metrics
\cite{DBLP:conf/hpca/JosephVT06},
\cite{DBLP:conf/micro/JosephVT06},
\cite{DBLP:conf/asplos/LeeB06}.
However, these endeavors only focus on the performance prediction of basic blocks and relative simple simulators rather than the complete benchmark programs and cycle-level simulators like gem5 \cite{binkert2011gem5}.
This stems from the intrinsic complexity and considerable computational overhead required to accurately simulate complete program and cycle-level processor behavior.

Another method to accelerate simulation is to run benchmark programs through sampling \cite{DBLP:journals/access/AkramS19} rather than executing the entire program — because benchmarks in SPEC 2017 suite often have billions of instructions to execute. It is really time-consuming and unnecessary to simulate the whole program to get the performance prediction results. In this approach, samples are sets of instructions that are deemed to be representative of the entire benchmark program. With sampling, rather than simulating the complete benchmark program, only a select number of representative instruction sets are simulated. Two ways are commonly used when picking out the correct sampling sets: statistical sampling and targeted sampling \cite{DBLP:conf/asplos/SherwoodPHC02}. Statistical sampling involves randomly selecting samples from the entire instruction sequence, or picking samples at regular intervals. An example of this approach is the periodic sampling utilized by SMARTS \cite{DBLP:conf/isca/WunderlichWFH03}. Targeted sampling involves selecting sampling points after investigating the behavior of benchmark. SimPoint \cite{DBLP:conf/sigmetrics/PerelmanHBSC03} is a targeted sampling tool that uses the number of times basic blocks are entered as a metric to detect program behavior. While sampling technologies significantly reduce the number of instructions that need to be sequentially simulated, there remains a time-consuming process due to the millions of instructions in one interval that still require accurate simulation.

To address the above issues, we propose a novel CPU simulator called CAPSim, which combines a fast functional instruction simulator and a machine-learning-based perfromance predictor. The CAPsim is capable of handling a complete benchmark by obtaining the instruction trace from function simulation and sum up the execution time of every small fragment of the trace (much smaller than the checkpoints of simpoint). 
The performance predictor using an attention-based neural network to predict the running time of each code fragment by disclosing the relationship of instructions in a trace that is key to the CPU performance. This segmentation exploits the parallel processing of smaller code pieces, leading to a significant enhancement in prediction speed. To improve accuracy, we have incorporated a context matrix into the code segments, which comprises architectural register values.
 
In conclusion, this paper presents the following contributions:

\begin{itemize}
\item We propose a CPU performance prediction method based on the attention mechanism. By leveraging the attention mechanism to uncover the relationships among instructions, we can quickly predict the execution cycles of instruction interval. This approach extends the prediction capability from the basic block level to full program, including complex control flows and memory accesses, thereby enabling fast and accurate simulation of all programs such as SPEC2017~\cite{SPEC2017}.
\item We propose CAPSim, a simulator that integrates fast function simulator with fine-grained code blocks performance prediction, thereby achieving a balance between simulation speed and accuracy. In the process of developing CAPSim, we introduce an accelerated training method. By clustering and applying sample techniques, we partition the training dataset into major categories such as compute-intensive, memory-intensive, and control-intensive. This method results in reduced training time, enhanced performance, and prevention of overfitting.
\item Our evaluation shows that CAPSim significantly accelerates benchmark simulation while maintaining accuracy comparable to state-of-the-art simulationbased predictors. As a result, we are able to achieve a maximum speedup of $7.78\times$ times during signal benchmark evaluation.
\end{itemize}

\section{Background and Motivation}

\subsection{Simulators}\label{Simulators}
Simulators have become indispensable tools in the realm of computer architecture research. Through sophisticated modeling, they facilitate the exploration of computer system behavior, enable accurate estimation of resource allocations, provide insights into program execution dynamics, and support the development of innovative architectural designs.  Many different kinds of simulators are playing different roles nowadays.

Several simulators such as llvm-mca \cite{llvmmca}, uiCA \cite{abel2022uica}, IACA \cite{Intel2017}, and OSACA \cite{laukemann2019automatic}, \cite{laukemann2018automated} can be used to predict the throughput of basic blocks, making them useful under certain circumstances such as compilation. Other kinds of simulators such as SimpleScalar \cite{austin2002simplescalar}, sniper \cite{DBLP:conf/sc/CarlsonHE11}, Zesto \cite{loh2009zesto}, MARSS \cite{DBLP:conf/dac/PatelACG11}, and gem5 \cite{binkert2011gem5} can model the execution of the whole program. 

Consider gem5 \cite{binkert2011gem5}, a cycle-level simulator which can be used to evaluate the running time of an O3  CPU on complex benchmark suite such as SPEC 2017.  By leveraging the customizable features of gem5, researchers have the flexibility to construct their own CPU architecture simulator, simulating the interactions between different hardware components including the memory system. As a result of the intricacies involved in its simulation, the cycle-level simulator has been experiencing challenges with efficiency during evaluation.

\subsection{Performance Predition and Multi-Head Attention Mechanism}\label{MHA}
Performance simulators differ fundamentally from functional simulators in that they must capture the running time of each instruction under various context information conditions. These conditions include data dependency, resource conflicts, and data preparation. Each of these factors can introduce variations in delay, leading to differences in the execution time. For example, if the same data is accessed in a preceding code segment, a subsequent code sequence that frequently accesses this data may execute faster because the data is already in the cache. Conversely, in an environment where the data is not preloaded in the cache, the execution time of the same code sequence would be adversely affected.
The execution order of instructions also plays a crucial role in overall performance. For example, consider a code sequence organized as: first load data, then perform an addition, and finally store the result. In this sequence, the pipelining mechanism can be fully leveraged. In contrast, if the sequence is reordered as: load data, then store data, then perform an addition, the STORE operation may block subsequent instructions, thereby increasing the overall execution time.

The use of transformers has been proven to be a great success in the field of natural language processing \cite{AttentionIsAllYouNeed}. The main technology in the transformer neural network lies in the encoder and decoder module called `Multi-Head Attention (MHA)', which is composed of an attention function. The attention function is defined as:

\begin{equation}
\begin{aligned}
\mathrm{Attention(\textbf{Q}, \textbf{K}, \textbf{V})=softmax(\frac{\textbf{QK}^T}{\sqrt{d}})\textbf{V}}
\label{eq_Attention}
\end{aligned}
\end{equation}
where the $\textbf{Q}, \textbf{K}, \textbf{V}$ are the inputs matrices of the attention function, called $query, key$ and $value$, respectively. $d$ is the dimension of input matrices $query$ and $key$. Then the MHA mechanism can be given by: 

\begin{equation}
\begin{aligned}
& \mathrm{MHA(\textbf{Q}, \textbf{K}, \textbf{V})=Concat(\textbf{H}_1, \ldots, \textbf{H}_h)\textbf{W}^{(O)}} \\
& \mathrm{where\ \textbf{H}_i=Attention(\textbf{QW}^{(Q)}_i, \textbf{KW}^{(K)}_i, \textbf{VW}^{(V)}_i)}
\label{eq_MHA}
\end{aligned}
\end{equation}
where $\textbf{W}^{(Q)}_i, \textbf{W}^{(K)}_i, \textbf{W}^{(V)}_i, \textbf{W}^{(O)}$ are parameter matrices which can be determined during training. 

The MHA mechanism proves to be more effective \cite{DBLP:journals/corr/abs-1909-06317} in handling extensive input text data than the Long Short-Term Memory (LSTM) network \cite{DBLP:journals/neco/HochreiterS97}, aiding researchers in achieving superior performance and accuracy. Similarities can be drawn between code blocks containing assembly instructions and natural language text data, as both are composed of basic words and require sequential processing. 
When performing performance simulation, the execution time of each instruction is computed and subsequently aggregated. Different code sequences, influenced by varying context information, yield differing execution times, rendering them particularly suitable to processing via attention mechanisms.

\subsection{Related Work}\label{RW}
Many works have explored the use of ML-based techniques in processors' performance prediction, as the traditional cycle-level simulators now require increasing simulation time and computational resources. In order to predict the CPI of superscalar processors, some of the works used crucial micro-architectural parameters to build linear regression models \cite{DBLP:conf/hpca/JosephVT06} or non-linear regression models \cite{DBLP:conf/micro/JosephVT06}. Despite CPI, Lee and Brooks \cite{DBLP:conf/asplos/LeeB06} predicted power by using regression modeling with cubic splines as well. Simple Regression model is a fundamental ML technique \cite{bonaccorso2018machine}. Regression model is a traditional machine learning technology, although it is effective, it may suffer from inaccuracy and lack of generalization in the face of more complex systems and programs.

Motivated by recent advancements in Deep Neural Networks (DNNs), several deep learning (DL) approaches have emerged to address the challenge of performance prediction. For instance, Andrew  \cite{adams2019learning} introduced a DL-based method aimed at predicting the performance of Halide programs for optimization purposes, albeit it necessitates intricate program features. Ithemal \cite{DBLP:conf/icml/MendisRAC19}, on the other hand, capitalizes on LSTM networks to predict the throughput of basic blocks, surpassing the accuracy of both mechanistic and empirical predictive models. Drawing inspiration from Ithemal, Difftune \cite{DBLP:conf/micro/RendaCMC20} adopts a similar framework and creates differentiable surrogates to fine-tune the parameters utilized in llvm-mca simulators. While these parameters enhance the predictive accuracy of llvm-mca on the test dataset, their scope remains confined to basic block throughput prediction rather than whole program analysis, employing a rudimentary LLVM simulator instead of a cycle-level simulator. In a related effort, Riyadh \cite{DBLP:conf/mlsys/BaghdadiMLAABA21} devised a deep learning based cost model for code optimization within the compiler. Facile \cite{DBLP:conf/iiswc/AbelSR23} serves as an interpretable predictor for basic block throughput, emphasizing the analytical aspect of simulation. Meanwhile, Granite \cite{DBLP:conf/iiswc/SykoraPMY22} offers a fresh perspective on basic block throughput prediction by integrating graph neural networks into the predictors.

\subsection{Motivation}\label{NM}
As it stands, existing research has been limited in scope, focusing either on predicting basic block throughput rather than analyzing entire programs, or utilizing simplistic simulators typically found in compilers rather than the cycle-level simulators commonly employed in computer architecture studies. To overcome these shortcomings, we introduce a novel simulator CAPSim, designed to simulate benchmark programs fast and accurately. 

Through the implementation of attention-based neural network performance predictor and the utilization of an advanced code trace clip extraction method, we have successfully achieved remarkable acceleration in benchmark evaluations, while maintaining accuracy compared with other advanced researches.

\section{CAPSim architecture overview}
The CAPSim is an end-to-end comprehensive benchmark performance simulator that employs an attention-based neural network predictor. It is primarily designed for evaluating the performance of a superscalar CPU across various benchmarks.

\subsection{Incorporating Attention Mechanism in Performance Prediction}
The primary innovation of CAPSim is the construction of a network that leverages the attention mechanism to reveal the relationships within instruction intervals.

Assume that the code sequence includes N instructions, and the total execution time is the sum of the execution times of individual instructions. Furthermore, each instruction's execution time depends on its ideal execution time as well as its related context information. Thus, we have:
\begin{equation}
\label{eq_1}
T_{total} = \sum_{n = 1}^{N} t_{i} \cdot \alpha_{i}
\end{equation}
where $t_{i}$ denotes the ideal execution time of the i-th instruction. In real execution situation, the actual execution time of an instruction is influenced by a multitude of factors, represented as $\alpha_{i}$. At the processor front-end, issues such as I-cache misses and branch mispredictions can deteriorate performance, while at the processor back-end, factors including D-cache miss, high latencies in functional units, and memory access failures can negatively impact execution time~\cite{10.1145/1168919.1168880}.
Therefore, the effects of these various factors on the instruction execution time can be transformed into a projection operation between high-dimensional vectors. Suppose that $F_{i}(x) = \mathbf{W}_{i} \boldsymbol{x} + \boldsymbol{b}_{i}$, which represents a linear transformation applied to vector $\boldsymbol{x}$, then
\begin{equation}
\begin{aligned}
\label{eq_2}
T_{total} &= F_{p}(...\sigma_{2}F_{2}(\sigma_{1}F_{1}(\mathbf{T} \times \mathbf{A}))...) \\
          &= MLP\Biggl(
            \left[
            \begin{array}{lll}
            t_{0,0}  & \cdots & t_{N-1,0}\\
            t_{0,1}  & \cdots & t_{N-1,1}\\
            \vdots  & \ddots & \vdots              \\
            t_{0,E-1}  & \cdots & t_{N-1,E-1}
            \end{array}
            \right] \\
            &\times
            [\alpha_{0}, \alpha_{1}, \ldots, \alpha_{N-1}]^{T}
            \Biggr)
\end{aligned}
\end{equation}
In this equation, vectors (rather than scalars) represent the ideal execution time for each instruction, effectively capturing the multifactorial effects influencing the actual execution time. For example, vector $\boldsymbol{t}_{0} = \{t_{0,0}, t_{0,1}, ..., t_{0,E-1}\}$ represents the ideal execution time of the first instruction. $\sigma$ represents a non-linear activation function, $E$ denotes the length of the vectors, and MLP represents a multilayer perceptron (MLP) network that maps vectors into a scalar with $p$ layers.

To derive the ideal execution time vector for each instruction, we employ an attention mechanism to analyze the instructions. Consider the i-th instruction,
\begin{equation}
\boldsymbol{C}_{i} = \{token_{0}, token_{1}, \ldots, token_{L-1}\}
\end{equation}
comprising $L$ tokens. Each token is mapped to an $E$-dimensional vector, so that the i-th instruction is represented as
\begin{equation}
\mathbf{CM}_{i} = \{\boldsymbol{emb}_{0}, \boldsymbol{emb}_{1}, \ldots, \boldsymbol{emb}_{L-1}\}
\end{equation}
where $\mathbf{CM}$ is an $(N \times L \times E)$ matrix. Applying the self-attention mechanism to $\mathbf{CM}_{i}$ produces
\begin{equation}
\label{eq_3}
\mathbf{RM}_{i} = Attention(\mathbf{CM}_{i}, \mathbf{CM}_{i}, \mathbf{CM}_{i})
\end{equation}
where $\mathbf{RM}_{i}$ is the resultant matrix that correlates with the ideal execution time of the i-th instruction. The token design in $\boldsymbol{C}_{i}$ ensures that the first vector of $\mathbf{RM}_{i}$ serves as the ideal execution time vector, denoted as $\boldsymbol{RT}_{i}$, with a shape of $(1 \times E)$. By processing all N instructions with self-attention, we obtain N ideal execution time vectors, which are subsequently stacked to form the ideal execution time matrix $\mathbf{T}_{E \times N}$. This can be expressed as
\begin{equation}
\begin{aligned}
\mathbf{T}_{E \times N} = \begin{pmatrix}\boldsymbol{RT}_{0}\\
                                         \boldsymbol{RT}_{1}, \\
                                         ..., \\
                                         \boldsymbol{RT}_{N-1}\end{pmatrix}^{T}
\end{aligned}                            
\end{equation}

Assuming each instruction is associated with $M$ context vectors, $\boldsymbol{context}_{i}$, that influence its execution behavior, the product of the ideal execution time matrix $\mathbf{T}_{E \times N}$ and the influence factor vector $\boldsymbol{\alpha}_{N \times 1}$ in Equation~\ref{eq_2} can be reformulated—through the introduction of trainable parameter matrices and normalization steps—as an attention relationship between the context matrix $\mathbf{contextM}_{M \times E}$ and $\mathbf{T}_{E \times N}$. Consequently, we obtain
\begin{equation}
\begin{aligned}
\label{eq_4}
T_{total} &= \sum_{i = 1}^{M} MLP\Bigl(Attention(\\
            & \mathbf{contextM}_{M \times E}, \mathbf{T}_{E \times N}^{T}, \mathbf{T}_{E \times N}^{T})\Bigr)
\end{aligned}
\end{equation}

Detailed performance predictor working flow and the construction of $\mathbf{contextM}_{M \times E}$ is provided in Section~\ref{PPD}.

\begin{figure}[t]
\centering
\includegraphics[width=0.99\linewidth]{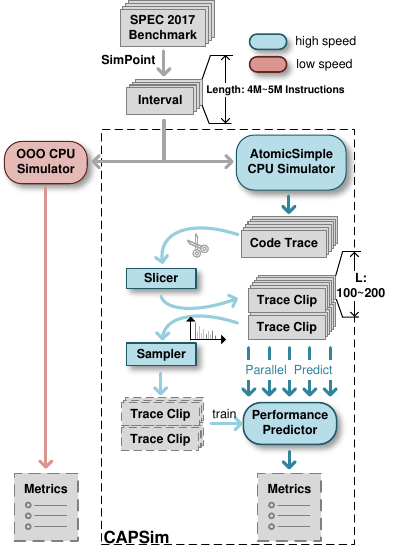}
\caption{The workflow of CAPSim.}
\label{fig_overview}
\end{figure}

\subsection{CAPSim Workflow}
The comparative study between cycle-level simulation using an O3 superscalar simulator and CAPSim is illustrated in Fig.~\ref{fig_overview}. First, programs from the SEPC 2017 benchmark are partitioned into several intervals using SimPoint. In the conventional approach, the O3 CPU simulator (developed with gem5 for the Power ISA) is employed to gather metrics for every interval. However, since each interval typically encompasses approximately one to two million cycles, this process becomes exceedingly time-consuming owing to the limited degree of parallelism. In contrast, as shown on the right, our CAPSim uses the AtomicSimple CPU simulator, which can simulate an interval at a significantly higher speed while providing the code trace of benchmark intervals for the predictor. This increased speed is attributable to its atomic operation model, where memory operations and instructions are executed in a single step, though it sacrifices timing precision.

Nevertheless, the substantial volume of code trace clips can considerably impede the training of the performance predictor. To accelerate training, a sampler selects a representative subset of the code trace clips, which substantially reduces the overall dataset used to train the predictor. Additionally, the remaining clips are categorized into several classes, and these categories are subsequently employed to further refine the training. This classification-based training method enables CAPSim to predict the execution time of previously unseen benchmarks both accurately and rapidly.

\section{Code Trace Clips Set Generation}\label{CodeSet}
\begin{figure}[t]
\centerline{\includegraphics[width=0.99\linewidth]{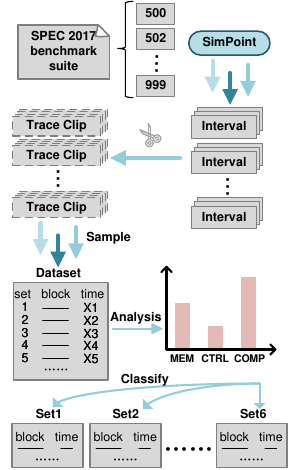}}
\caption{The workflow of producing code trace clips from the SPEC 2017 suite.}
\label{fig_dataset}
\end{figure}

Fig.~\ref{fig_dataset} depicts the overall flow of applying SimPoint, Instruction Slicer and Sampler to generate the code trace clips set for the CAPSim, along with the clip sets classification process. 
By following this flow, we can ensure that the sampled code trace clips are diverse and representative of the program's behavior, enabling the CAPSim to make accurate predictions with great speed and reasonable training time. Besides, based on observations of program behavior in SPEC2017, we categorized all benchmark programs using three labels: control-intensive, computation-intensive, and memory-intensive. Each benchmark can have multiple labels. Using these labels, we grouped all benchmarks in SPEC2017 into six distinct sets. The specific classification details will be described in Section \ref{MAE}.

\begin{figure}[t]
\centerline{\includegraphics[width=0.85\linewidth]{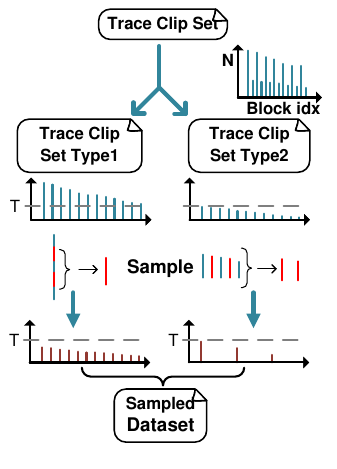}}
\caption{The workflow of the trace clip sampler.}
\label{fig_sampling}
\end{figure}

\begin{figure*}[!ht]
\centerline{\includegraphics[width=0.9\linewidth]{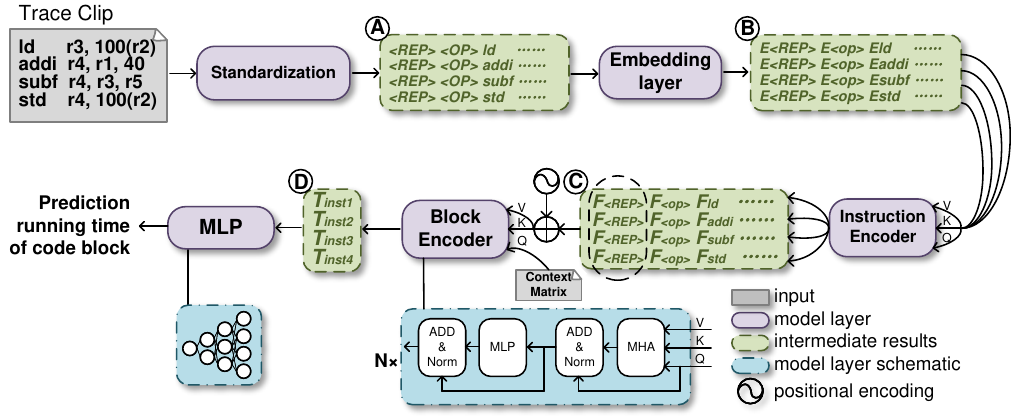}}
\caption{Overview of the performance predictor.}
\label{fig_network}
\end{figure*}


\begin{algorithm}[t]
    \caption{Generate code trace clips set from instruction sequence}
    \label{algorithm1}
    \KwIn{Set of the instruction trace sequence $I$}
    \KwOut{Set of the code trace clips $B$}
    $B \leftarrow \varnothing$\;
    $b \leftarrow \varnothing$\;
    $InstPrev \leftarrow I[0]$\;
    $BlockLength \leftarrow 0$\;
    $TimePrev \leftarrow 0$\;
    $TimeBegin \leftarrow 0$\;
    \ForEach{inst \textbf{in} $I$}
    {$InstNow \leftarrow inst$\;
    $TimeNow \leftarrow inst.CommitTime$\;
    $b.append(InstPrev)$\;
    $BlockLength \leftarrow BlockLength + 1$\;
    \If{$BlockLength \geq L_{min}\ \textbf{and}\newline\,TimeNow \neq TimePrev$}{
        $b.time \leftarrow TimePrev - TimeBegin$\;
        $B.append(b)$\;
        $TimeBegin \leftarrow TimePrev$\;
        $b \leftarrow \varnothing$\;
        $BlockLength \leftarrow 0$\;
        }
    $InstPrev \leftarrow InstNow$\;
    $TimePrev \leftarrow TimeNow$\;}
\end{algorithm}

\subsection{Instruction Sequence Slicer}\label{Slicer}
When the simulator restores the checkpoint, it recovers the saved values to the registers and memory system. After that, millions of instructions (dependent on the interval size) are going to be executed to get metrics of the simulation system. In our work, we implement an O3 superscalar processor simulator based on Power ISA using gem5.

The instruction sequence slicer has been developed to meticulously analyze all instructions within an interval, generating a comprehensive set of code trace clips that encompass the instructions along with their respective execution time. By utilizing this tool, the predictor is able to predict the overall performance by considering the performance of all code trace clips. Additionally, shorter trace clips can be processed by the neural network simultaneously, revealing potential parallelism within the given interval checkpoint.

Given the set of instruction sequence in an interval restoring process, Algorithm~\ref{algorithm1} outlines the procedure for generating a set of code trace clips from it. The identification of the starting and ending points of each code clip is guided by two key principles: Firstly, the clip must contain a number of instructions surpassing the predefined threshold $L_{min}$. Additionally, the commit time of the initial instruction within the clip must differ from that of the final instruction in the preceding clip. Adhering to these guidelines serves to prevent two potential issues: Firstly, the inclusion or exclusion of certain instructions within a code clip should not impact its overall runtime. Secondly, it ensures uniformity in the lengths of the code clips. Following the identification of the starting and ending instructions, the runtime of the clip is ascertained by the disparity in commit times between the initial instruction within the clip and the final instruction of the previous clip.

\subsection{Code Trace clips Sampler}\label{Sampler}
Following the instruction sequence slicing process, a collection of code trace clips is presented along with their execution times. With an interval size of 5,000,000 and a minimum code trace clip length threshold of $L_{min}$ is 100, approximately 50,000 code trace clips are contained within each interval. Considering that the SPEC 2017 suite contains 623 interval checkpoints, the total count of code trace clips reaches around 30,000,000 - a substantial load for model training process. In light of this, it is imperative to undertake additional sampling of the code trace clips set to alleviate the strain on the model training process.

In order to sample the generated code trace slips sets, firstly, we sort each trace clip with unique code sequence content. Then, after setting a threshold of occurrence, we can divide the clips into two categories. One category consists of a significant repetition of identical code trace clips (occurrence times bigger than threshold), while the other category features a smaller quantity of diverse unique code trace clips.

Following the sorting process, it is recommended to periodically sample from the two distinct types of code trace clips sets. In the case of code clips surpassing the occurrence threshold, it is advised to sample them individually within their respective categories. This approach involves lowering the occurrence number of each code trace clips while ensuring the preservation of category distribution. Conversely, for code clips falling below the specified threshold, it is suggested to sample them across categories, resulting in a reduction of categories represented for these code trace clips instead of adjusting their occurrence number. The process of sampling is shown in Fig.~\ref{fig_sampling}.

\section{Performance Predictor Design}\label{PPD}
\begin{figure*}[t]
\centering
\subfigure[Standardization of ``addi'']{\includegraphics[width=0.32\textwidth]{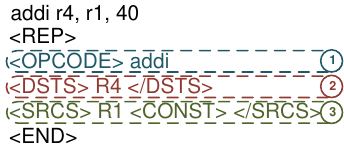}}
\subfigure[Standardization of ``ld'']{\includegraphics[width=0.32\textwidth]{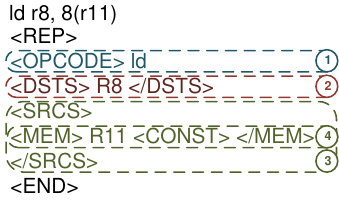}}
\subfigure[Standardization of ``cmpi'']{\includegraphics[width=0.32\textwidth]{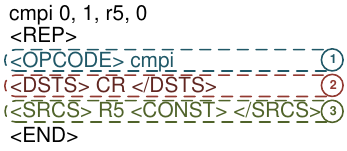}}
\caption{Examples of standardization transformation.}
\label{fig_asm}
\end{figure*}

\begin{figure}[h!]
\centering
\subfigure[Example of processing register ``R10'' into Register Matrix]{\includegraphics[width=0.40\textwidth]{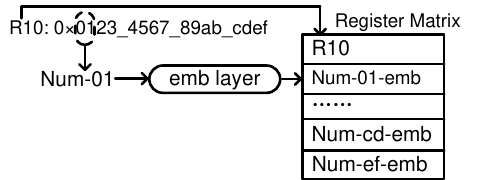}}
\subfigure[Process of producing Context Matrix]{\includegraphics[width=0.4\textwidth]{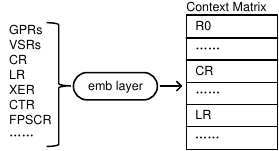}}
\caption{Construction of Context Matrix.}
\label{fig_contextmatrix}
\end{figure}

Fig.~\ref{fig_network} outlines the overarching framework of the performance predictor. By partitioning the entire instruction trace sequences into several trace clips, the predictor is able to utilize these clips as input and generate an end-to-end prediction about the execution time of the code trace.
The performance predictor can be deconstructed into the subsequent components: standardization transformation, token embedding layer, instruction encoder, block encoder, and MLP layer.
\subsection{Standardization Transformation}\label{SL}
\begin{table*}[t]
\caption{Registers used in Context Matrix in Power ISA}
\begin{center}
\begin{tabular}{lccl}
\toprule[1pt]
\textbf{Registers} & \textbf{Number} &\textbf{Valid Width} & \textbf{Description} \\ 
\hline
General Purpose Registers (GPR)          & 32    & 64    & GPR is the principal storage registers. \\
Vector-Scalar Registers (VSR)            & 64    & 128   & VSR is used to perform floating-point computations. \\
Floating-Point Status and Control Register (FPSCR)    & 1     & 32    & FPSCR controls the handling of floating-point exceptions. \\
Condition Register (CR)                  & 1     & 32    & CR reflects the result of certain operations. \\
Vector Status and Control Register (VSCR)& 1     & 32    & VSCR controls the handling of vector opreation exceptions. \\
Current Instruction Address (CIA)        & 1     & 64    & CIA saves the current instruction address value. \\
Next Instruction Address (NIA)           & 1     & 64    & NIA saves the next instruction address value. \\
Link Register (LR)                       & 1     & 64    & LR can be used to provide the branch target address. \\
Fixed-Point Exception Register (XER)     & 1     & 64    & XER saves exceptions signals in fixed-points manipulation. \\
Count Register (CTR)                     & 1     & 64    & CTR can be used to hold a loop count. \\
\hline
\end{tabular}
\label{table_context}
\end{center}
\end{table*}

The standardization transformation transforms individual raw assembly instructions within the trace clip into a structured format composed of standardized tokens. Subsequently, the token embedding layer is able to transform each token into the embedding vector \cite{mikolov2013distributed}.

Our design involves the organization of the standardization format into four segments, excluding the tokens \mbox{\textless REP\textgreater} and \mbox{\textless END\textgreater}. The first segment is denoted by the token \mbox{\textless OPCODE\textgreater}, encompassing the operational code symbol of the assembly instruction. The second segment, enclosed by the tokens \mbox{\textless DSTS\textgreater} and \mbox{\textless /DSTS\textgreater}, denotes the destination operand symbol of the instruction. Following this, the third segment denotes the source operand symbol in the instruction, discerned by the tokens \mbox{\textless SRCS\textgreater} and \mbox{\textless /SRCS\textgreater}. Furthermore, to point out the memory access behavior, we introduce the fourth segment delineated by \mbox{\textless MEM\textgreater} and \mbox{\textless /MEM\textgreater}.

It is important to emphasize the configurability of all four segments, as certain instructions may not require memory access and there are control flow instructions, such as jump instructions, that do not necessitate the use of source operands.

We present three instances demonstrating the transformation of raw assembly instructions in Power ISA into a standardized format. In Fig.~\ref{fig_asm}(a), we illustrate a common scenario where a constant number is transformed into the token \textless CONST\textgreater. Fig.~\ref{fig_asm}(b) provides an illustration of a load instruction utilizing memory as the source operand. Fig.~\ref{fig_asm}(c) showcases a slightly perplexing example where the destination operand is not explicitly stated in the instruction $cmpi$. In Power ISA, certain control registers such as the condition register (CR), link register (LR), and program counter register (CIA) are not explicitly included in the assembly instruction, yet they hold significant importance in the flow of execution. Therefore, it is imperative to manually incorporate them during the standardization process.

\subsection{Context Information}\label{CI}
The predictor leverages the context information of code trace clips to enhance the accuracy of prediction results. In our design, context information refers to the CPU state before executing the trace clip, including various kinds of register values.

The illustration shown in Fig.~\ref{fig_contextmatrix}(a) demonstrates the process of transforming a signal register along with its corresponding value into a component of the context matrix referred to as the 'Register Matrix'. Let's consider the example of a general-purpose register (GPR) denoted as R10, with a value of 0x0123\_4567\_89ab\_cdef. In the beginning, the output of the standardization layer includes tokens designated with the register's name, such as GPR and control registers. The initial entry in the context matrix represents the embedding vector of R10, which is the same with standardization layer's result. Subsequently, the register's value is segmented into 16 groups based on each two of hexadecimal numbers. Each group forms a distinct token that will be converted into an embedding vector. Finally, all embedding vectors are concatenated to construct the context matrix.
Assume that we have $M$ embedding vectors, denoted by $\boldsymbol{context}_{0}, \boldsymbol{context}_{1}, \ldots, \boldsymbol{context}_{M-1}$, then the context matrix can be expressed as
\begin{equation}
\begin{aligned}
\mathbf{contextM}_{M \times E} = \begin{pmatrix}\boldsymbol{context}_{0}\\
                                                \boldsymbol{context}_{1}\\
                                                \ldots\\
                                                \boldsymbol{context}_{M-1}\end{pmatrix}
\end{aligned}
\end{equation}

By performing such a transformation to registers in the context, we can obtain the context matrix for each trace clip in every prediction. This process is illustrated in Fig.\ref{fig_contextmatrix}(b). The specific registers utilized to establish the context information can be found in Table~\ref{table_context}. When designing the gem5 model in Power ISA, Vector-Scalar Registers are used to realize the function of Floating-Point Registers.

\begin{figure*}[th!]
\centering
\includegraphics[width=0.95\linewidth]{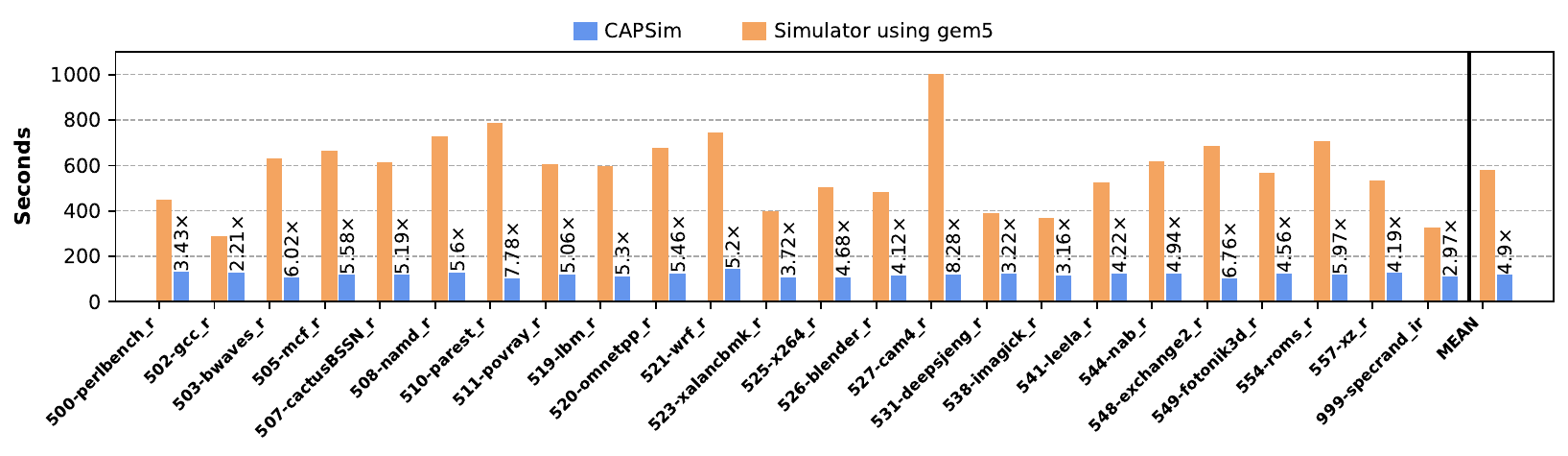}
\caption{Speed comparison between simulator on CPU and predictor on GPU.}
\label{fig_speed}
\end{figure*}

\subsection{Prediction Flow}\label{PF}
In Fig.~\ref{fig_network}, considering the input code trace clip with a length of $\textbf{L}_{clip}$, and the maximum token count after standardization is $\textbf{L}_{token}$, resulting in the intermediate result A matrix being of a shape equal to $(\textbf{L}_{clip}, \textbf{L}_{token})$. Subsequent to the embedding layer, each token within the matrix will be substituted with a corresponding embedding vector, whose dimension is E. Consequently, the shape of the intermediate result B will be $(\textbf{L}_{clip}, \textbf{L}_{token}, \textbf{E})$. In result B, every matrix possessing a shape of $(\textbf{L}_{token}, \textbf{E})$ serves as the result of a standardization layer and embedding layer, and within this block, there exist $\textbf{L}_{clip}$ instructions.

Each instruction embedding matrix undergoes processing by the instruction encoder layer, implementing a self-attention mechanism in a sequential manner. The third intermediate outcome is generated through stacking all results from the instruction encoder layer, maintaining a shape of $(\textbf{L}_{clip}, \textbf{L}_{token}, \textbf{E})$. To enable successful processing of the third intermediate result by the block encoder layer, the embedding vector of token \mbox{\textless REP\textgreater} serves a crucial role. The additional \mbox{\textless REP\textgreater} token embedding acts as a learnable embedding \cite{dosovitskiy2020image} to represent the entire single instruction, with each \mbox{\textless REP\textgreater} embedding forming the input matrix for the block encoder layer. To account for the instruction sequence within the trace clip, positional encoding is also added to the input matrix prior to the layer. The block encoder executes an attention mechanism between context information and the third intermediate result, resulting in a fourth output with a shape mirroring that of the context matrix, denoted as $(\textbf{L}_{context}, \textbf{E})$. The final layer of the predictor consists of an MLP with an arithmetic mean function, which transforms the last intermediate result into the final prediction for the execution time of the trace clip.

\begin{figure}[!ht]
\centering
\subfigure[Original distribution of code trace clips]{\includegraphics[width=0.49\textwidth]{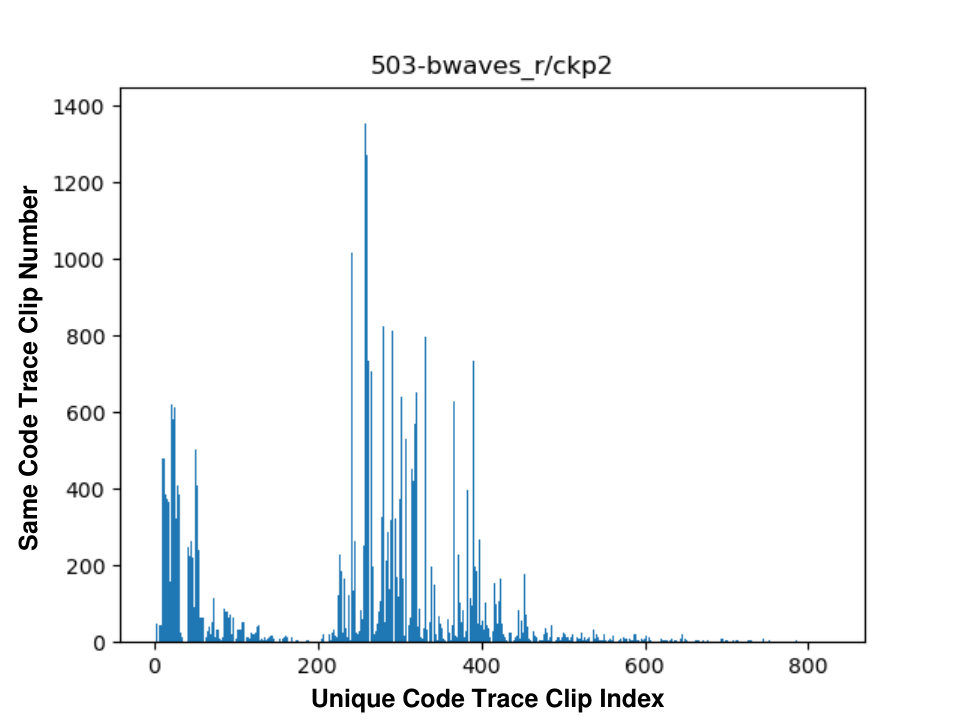}}
\subfigure[Sorted distribution of code trace clips]{\includegraphics[width=0.49\textwidth]{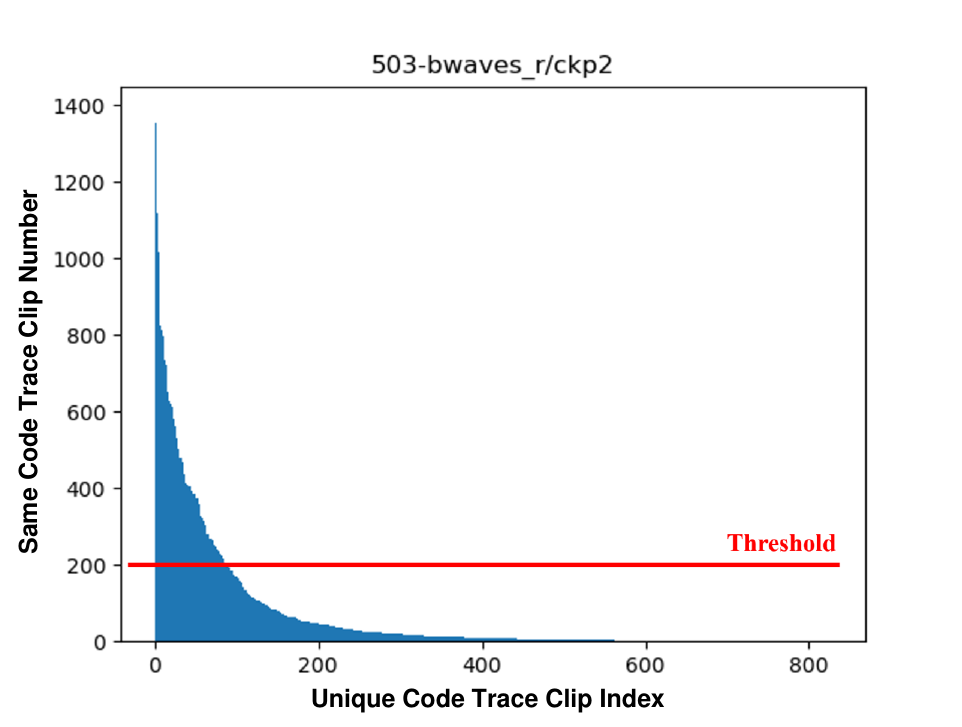}}
\caption{The distribution of code trace clips in the second interval of benchmark 503-bwaves\_r.}
\label{fig_502ckp3_ab}
\end{figure}

\begin{table}[t]
\caption{Checkpoint Numbers of Each Benchmark in SPEC 2017 suite}
\begin{center}
\begin{tabular}{cccc}
\toprule[1pt]
\textbf{Name} & \textbf{CKP Num} & \textbf{Tag} & \textbf{Set No.} \\ 
\hline
500.perlbench	&   7	& CTRL      & 1\\
502.gcc	        &   1	& CTRL      & 2\\
503.bwaves	    &   24	& COMP+MEM  & 1\\
505.mcf	        &   32	& COMP+MEM  & 2\\
507.cactuBSSN	&   20	& COMP+MEM  & 3\\
508.namd	    &   70	& COMP+MEM  & 4\\
510.parest	    &   78	& COMP+MEM  & 5\\
511.povray	    &   16	& COMP+MEM  & 6\\
519.lbm	        &   16	& COMP+MEM  & 1\\
520.omnetpp	    &   26	& CTRL      & 3\\
521.wrf	        &   71	& COMP+MEM  & 2\\
523.xalancbmk	&   5	& CTRL+MEM  & 4\\
525.x264	    &   13  & COMP      & 3\\
526.blender	    &   13  & COMP+MEM  & 4\\
527.cam4	    &   86  & COMP+MEM  & 5\\
531.deepsjeng	&   4   & CTRL      & 5\\
538.imagick	    &   4   & COMP+MEM  & 6\\
541.leela	    &   11  & CTRL+MEM  & 1\\
544.nab	        &   17  & COMP+MEM  & 2\\
548.exchange2	&   40  & CTRL+MEM  & 6\\
549.fotonik3d	&   15  & COMP+MEM  & 3\\
554.roms	    &   43  & COMP+MEM  & 4\\
557.xz	        &   8   & COMP+MEM  & 5\\
999.specrand	&   3   & COMP+MEM  & 6\\
\hline
\end{tabular}
\label{table_ckpnums}
\end{center}
\end{table}



\section{Evaluation}
This section reports and analyzes the outcomes of using the predictor to estimate the runtime of SPEC 2017 benchmarks. The golden results were generated by a Power ISA out-of-order superscalar simulator built with gem5. We start by describing the experimental setup and dataset parameters. Next, we outline the implementation of the predictor model, including its parameters, training, validation, and testing processes. We then showcase the predictor’s speed improvements over the traditional gem5 simulator. Finally, we evaluate the prediction accuracy across different scenarios and simulator parameter configurations.

\subsection{Setup}\label{Setup}
Our predictor model is trained and evaluated using the NVIDIA GeForce RTX 4090 with a memory capacity of 24 GB.
The gem5 simulator is designed to simulate an IBM Power8 CPU and runs on an Intel Xeon CPU E5-2623 v4. Released in Q1 2016, this processor achieves a single-core score of 935 and a multi-core score of 3442 (using 4 cores)~\cite{Intel_Xeon_1}. Meanwhile, the Intel Xeon Platinum 8380—launched in Q2 2021—delivers a single-core score of 1239 and a multi-core score of 3980 (using 4 cores)~\cite{Intel_Xeon_2}. Although the E5-2623 v4 is nearly five years older than the Platinum 8380, its performance disadvantage is not obvious. Therefore, employing the Intel Xeon CPU E5-2623 v4 in this work does not result in an unfair comparison with the RTX 4090.

In the construction of our code trace clips, an interval size of 5,000,000 is employed, alongside a warm-up size equal to 1,000,000. The classified information of each benchmark and detailed checkpoint numbers in the SPEC2017 benchmark suite are provided in Table~\ref{table_ckpnums}. The sampler incorporates a threshold of 200 and a sampling coefficient of 0.02. Under this configuration, the training time will be reduced from 300 hours to about 10 hours.

\begin{figure}[t!]
\centerline{\includegraphics[width=0.99\linewidth]{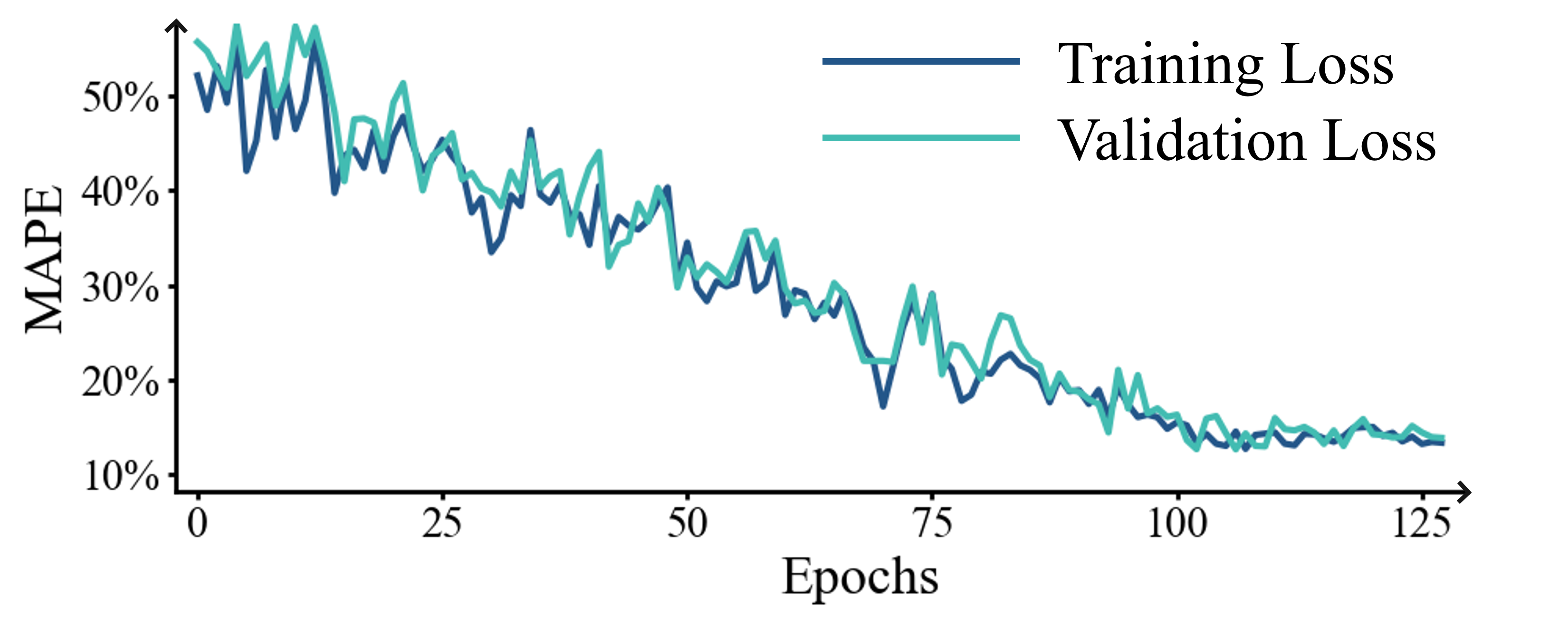}}
\caption{Training Loss vs. Validation Loss.}
\label{fig_tl}
\end{figure}

\subsection{Model Implementation and Training}\label{MIT}
\begin{figure*}[h!]
\centerline{\includegraphics[width=0.99\linewidth]{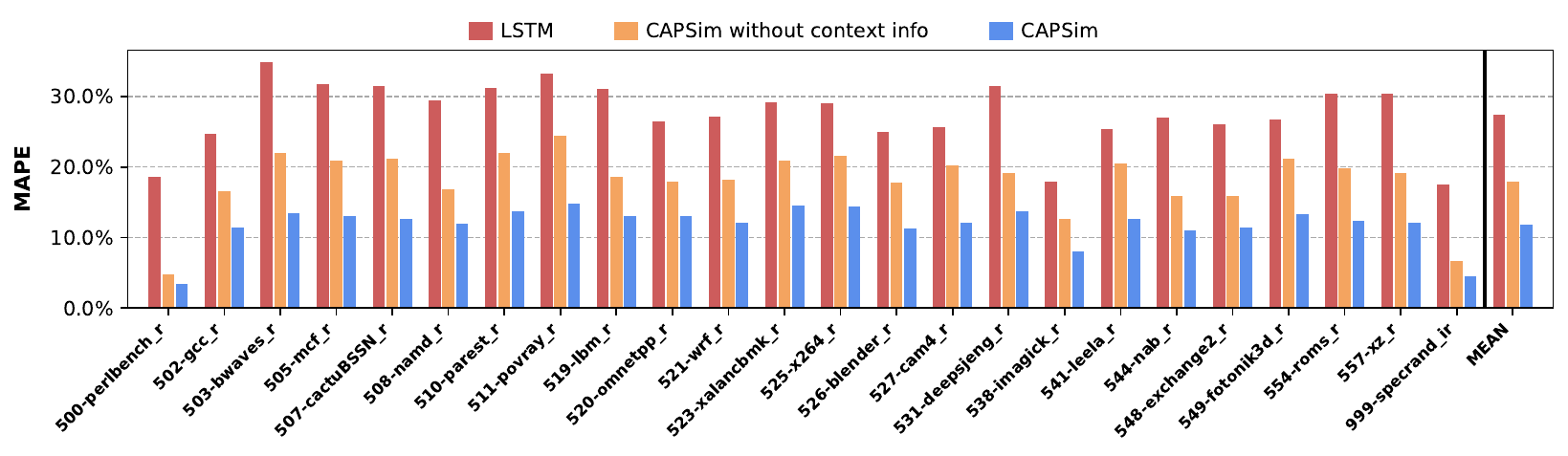}}
\caption{Average error of different predictors.}
\label{fig_acc}
\end{figure*}

The model we evaluate has an embedding vector size of 128. The number of heads in MHA is four, and the layer used in the instruction encoder and block encoder is four.We conduct the training and testing process using two different approaches. The first approach involves mixing all the code trace clips from 24 benchmarks, assigning 80\% of the generated code clips dataset to the training set, 10\% to the validation set, and 10\% to the test set. The loss function we use is Mean Absolute Percentage Error (MAPE). We then test the accuracy of the trained model on all the benchmarks from SPEC2017. The second approach involves classifying the generated code clips according to their type (discussed in section~\ref{CodeSet}.) into six sets. We then perform training on one set and test the model’s accuracy on another set. This procedure demonstrates the accuracy of CAPSim on unseen benchmarks, which is a common scenario when using the simulator in architecture research. The detailed sets information is shown in Table~\ref{table_ckpnums}, in which CTRL, COMP, and MEM represent controlintensive, computation-intensive, and memory-access-intensive tags, respectively.

Fig.~\ref{fig_502ckp3_ab}(a) displays the distribution of code clips for checkpoint three in benchmark 503-bwaves\_r. Each code clip with unique instruction sequence content is represented on the x-axis in the order of appearance, while the y-axis shows the total number of each unique kind of code clip. After setting a threshold of occurrence, it is evident from the visualization that code clip fall into two categories within each interval. One category consists of a significant repetition of identical code clips, while the other category features a smaller quantity of diverse unique code clips. This phenomenon is more vivid in Fig.~\ref{fig_502ckp3_ab}(b), where the distribution of code clips is arranged in descending order based on the number of occurrence.

\begin{equation}
\begin{aligned}
\mathrm{\textbf{Loss}(prediction, fact) = \frac{\left\lvert prediction - fact \right\rvert }{fact}}
\label{eq_AMAPE}
\end{aligned}
\end{equation}

The trainer we use is Stochastic Gradient Descent (SGD), with an initial learning rate of 0.001 and momentum of 0.9. The training process stops at around epoch 128, as shown in Fig.~\ref{fig_tl}.

\subsection{CAPSim Inference Speed Evaluation}\label{MISE}
Fig.~\ref{fig_speed} shows the restoring time of each benchmark along with the inference time on the GPU, whereby the y-axis denotes the processing seconds of each benchmark. From the observations made on this figure, it is evident that the CAPSim has managed to achieve a notable speedup of up to 8.3× when compared with the restoration process carried out with the gem5. The arithmetic mean of speedup is 4.9×.

We can infer from the figure that CAPSim provides more speedup when evaluating a benchmark with a larger number of checkpoints compared to gem5. This is because when restoring checkpoints in gem5, it is typically done with a fixed level of parallelism (determined by the number of CPU cores). This leads to inefficiencies when the number of checkpoints exceeds the maximum number of processes, limiting the full parallelization of checkpoint evaluation. However,CAPSim running on GPUs can fully exploit the parallelism between checkpoints, thus enhancing execution speed.

\subsection{CAPSim Accuracy Evaluation}\label{MAE}
\begin{table}[t]
\caption{Average Error With Different Simulator Parameters}
\begin{center}
\begin{tabular}{ccccc}
\toprule[1pt]
\textbf{FetchWidth} & \textbf{IssueWidth} & \textbf{CommitWidth} & \textbf{ROBEntry} & \textbf{Error} \\ 
\hline
8 & 8 & 8 & 192 & 12.0\% \\
4 & 8 & 8 & 192 & 12.2\% \\
8 & 4 & 8 & 192 & 12.9\% \\
8 & 8 & 4 & 192 & 12.5\% \\
8 & 8 & 8 & 128 & 12.8\% \\
\hline
\end{tabular}
\label{table_acc}
\end{center}
\end{table}
    
To evaluate the accuracy of CAPSim, we first assessed the model using Method 1 as mentioned in Section~\ref{CodeSet}. In Fig.~\ref{fig_acc}, the attention mechanism’s advantage in handling longer code trace clips enables CAPSim achieved a 9.5\% to 21.2\% improvement in accuracy compared to LSTM-based model Ithemal, with an average accuracy increase of 15.8\%. This improvement underscores the effectiveness of the attention mechanism in code trace clips running time prediction.

To investigate the performance improvement brought by the introduction of context information, we also conducted an ablation study on CAPSim. Compared to the CAPSim without context information, CAPSim achieved a 1.3\% to 9.6\% increase in accuracy, with an average accuracy improvement of 6.2\%, which is also shown in Fig.~\ref{fig_acc}. This demonstrates that context information plays a critical role in improving accuracy of CAPSim.

\begin{figure}[h!]
\centerline{\includegraphics[width=0.99\linewidth]{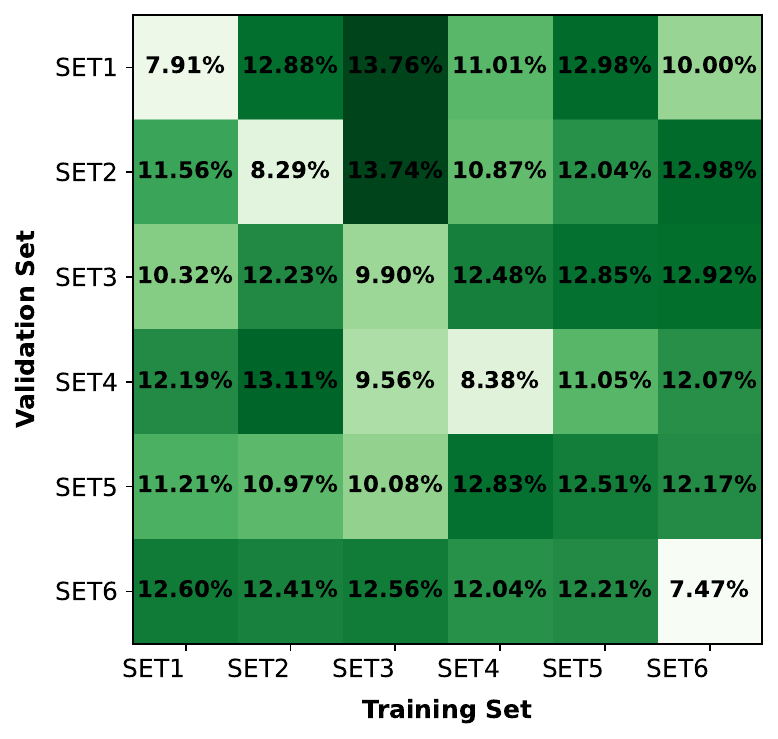}}
\caption{Accuracy results for 36 Train-Test combinations across six sets. Given in MAPE}
\label{fig_heatmap}
\end{figure}

When using simulators, it is necessary to evaluate the performance of the simulator across various benchmarks to explore the effects of certain microarchitecture. Method 2 in Section~\ref{CodeSet} explored CAPSim’s performance under such scenarios. We conducted 36 trainingevaluation tests on six sets of code clip collections composed of different benchmarks. Fig.~\ref{fig_heatmap} presents the accuracy results(given as MAPE) in these 36 test cases. For categorized benchmarks, CAPSim achieved an accuracy of 91.3\% on the training set, with an overall average accuracy of 88.3\%, representing a 0.5\% improvement compared to uncategorized datasets. This improvement is mainly due to the more comprehensive code trace coverage in the categorized training sets, demonstrating CAPSim’s generalization capability on unseen benchmarks.

Similarly, we tested CAPSim’s generalization capability for different microarchitectures by changing the microarchitecture parameters.
In conventional superscalar CPUs, various microarchitectural parameters critically influence overall performance. In this study, we investigate the effects of four key parameters: $FetchWidth$, $IssueWidth$, $CommitWidth$, and $ROBEntry$. To expedite the training process for CAPSim, we initially develop a basic version of the model. Subsequently, we modify one parameter at a time from this baseline to assess its generalization capability. Leveraging the pre-trained baseline reduces the network's initial error and accelerates training for configurations with different parameter settings. 

Table~\ref{table_acc} shows CAPSim’s accuracy performance under five different parameter configurations. The first one is treated as basic version. These selected changing parameters are considered fundamental in modern O3 CPU design, encompassing various hardware configurations within the processors. Although this accuracy level reflects a small trade-off when compared to specialized cycle-level simulators, it remains within an acceptable range for practical applications due to its significant speed advantage.

\section{Conclusion}

In our study, we introduce a predictor that utilizes an attention mechanism to expedite the evaluation of the running time of the complete benchmark within the SPEC 2017 suite using a cycle-level simulator. Our findings indicate that our predictor can achieve a speedup of up to $8.3\times $ when compared to a simulator operating on the CPU, while still preserving accuracy relative to alternative approaches. These results highlight the promising capabilities of employing cutting-edge deep learning neural networks in the realm of computer architecture.


\end{document}